\documentclass[onecolumn,secnumarabic,amssymb, nobibnotes, aps, pr,superscriptaddress]{revtex4-1}

\usepackage{amsmath,amssymb}
\usepackage{collectbox}
\usepackage{stackengine,graphicx}
\usepackage{bbm}
\usepackage{capt-of}

\def\l{\left}
\def\r{\right}

\newcommand{\f}{\frac}

\begin{document}

\title{How Cox models react to a study-specific confounder in a patient-level pooled dataset: Random-effects better cope with an imbalanced covariate across trials unless baseline hazards differ.}

\author{Thomas~McAndrew, Ph.D}
\email{tmcandrew@crf.org}
\affiliation{Cardiovascular Research Foundation}

\author{Bjorn~Redfors, M.D. Ph.D}
\affiliation{Cardiovascular Research Foundation}
\affiliation{Department of Cardiology Sahlgrenska University Hospital, Gothenburg, Sweden}

\author{Aaron~Crowley, M.A.}
\affiliation{Cardiovascular Research Foundation}

\author{Yiran~Zhang, M.S.}
\affiliation{Cardiovascular Research Foundation}

\author{Shmuel~Chen, M.D., Ph.D.}
\affiliation{Cardiovascular Research Foundation}
\affiliation{Hadassah Medical Center Jerusalem, Israel}

\author{Mordechai~Golomb, M.D.}
\affiliation{Cardiovascular Research Foundation}
\affiliation{Hadassah Medical Center Jerusalem, Israel}

\author{Maria~Alu, M.S.}
\affiliation{Cardiovascular Research Foundation}
\affiliation{Columbia University Medical Center}

\author{Dominic~Francese}
\affiliation{Cardiovascular Research Foundation}

\author{Ori~Ben-Yehuda, M.D.}
\affiliation{Cardiovascular Research Foundation}
\affiliation{Columbia University Medical Center}

\author{Akiko~Maehara, M.D.}
\affiliation{Cardiovascular Research Foundation}
\affiliation{Columbia University Medical Center}

\author{Gary~Mintz, M.D.}
\affiliation{Cardiovascular Research Foundation}

\author{Gregg~Stone, M.D.}
\affiliation{Cardiovascular Research Foundation}
\affiliation{Columbia University Medical Center}

\author{Paul~Jenkins, Ph.D.}
\affiliation{Bassett Research Institute}

\date{2018-05-07}

\begin{abstract}
  \textbf{Background:} Combining patient-level data from clinical trials can connect rare phenomena with clinical endpoints, but statistical techniques applied to a single trial may become problematical when trials are pooled. Estimating the hazard of a binary variable unevenly distributed across trials showcases a common pooled database issue.

  \textbf{Objectives:} We studied how an unevenly distributed binary variable can compromise the integrity of fixed and random effects Cox proportional hazards models.

  \textbf{Methods:} We compared fixed effect and random effects Cox proportional hazards models on a set of simulated datasets inspired by a 17-trial pooled database of patients presenting with ST-segment elevation myocardial infarction (STEMI) and non-STEMI undergoing percutaneous coronary intervention.

  \textbf{Results:} An unevenly distributed covariate can bias hazard ratio estimates, inflate standard errors, raise type I error, and reduce power.
  While uneveness causes problems for all Cox proportional hazards models, random effects suffer least.
  Compared to fixed effect models, random effects suffer lower bias and trade inflated type I errors for improved power.
  Contrasting hazard rates between trials prevent accurate estimates from both fixed and random effects models.

  \textbf{Conclusions:} When modeling a covariate unevenly distributed across pooled trials with similar baseline hazard rates, Cox proportional hazards models with a random trial effect more accurately estimate hazard ratios than fixed effects.
  Differing between-trial baseline hazard rates bias both random and fixed effect models.
  With an unevenly-distributed covariate and similar baseline hazard rates across trials, a random effects Cox proportional hazards model outperforms a fixed effect model, but cannot overcome contrasting baseline hazard rates.

\end{abstract}
\maketitle

\section{Introduction}

Pooling data from several clinical trials~\cite{lo2015sharing,mello2013preparing,ross2013ushering} can create robust results for endpoints too rare to study within any single trial~\cite{prospective2002age,hart2007meta,cannon2006meta,sjauw2009systematic,mamas2012influence,spaulding2007pooled,caixeta20095,steg2013effect}, but studying a covariate unequally distributed across trials (due to differences in eligibility criteria, definitions, or other study-specific factors) could lead to inaccurate conclusions~\cite{chalmers1991problems,flather1997strengths,berman2002meta}.
Cox proportional hazards (cph) models~\cite{cox1984analysis} associate endpoints with patient characteristics and control between-trial differences through stratification, or by including trial as a fixed or random effect.
Stratifying allows the per-trial baseline hazard rates to take any form, while both fixed and random effects models stiffen assumptions about how the hazard rate varies by trial.

Past studies compared the performance of these three types of models~\cite{andersen1999testing,localio2001adjustments,glidden2004modelling} by modifying a simulated trial-effect (i.e. differing numbers of trials, more varied trial baseline hazards), but previous studies have not considered modeling a binary covariate with only one level per clinical trial.
Studying an imbalanced covariate across pooled clinical trials will answer whether we can glean sensible statistical estimates from clinical trials with varied purposes.
We compared stratified, fixed, and random effects Cox proportional hazards model's ability to estimate an association between a simulated binary covariate unequally distributed across trials and clinical endpoints.

\section{Methods}

\subsection{Pooled DES study data}

We pooled data from $17$ coronary stent trials comparing drug-eluting stents (1st and 2nd generation) to bare-metal stents from $2006$ to $2013$ into a single dataset (pooled-DES).
This enabled us to study how the $26,564$ patient's clinical presentation (ST-segment elevation myocardial infarction [STEMI] versus non ST-segment elevation myocardial infarction [NSTEMI]) impacts mortality, myocardial infarction, bleeding, revascularization, and stent thrombosis at $5$ years while also adjusting for trial-specific differences in baseline hazard rates.

The majority of trials enrolled NSTEMI patients and only one pooled trial (HORIZONS-AMI) enrolled STEMI patients. This pooled dataset inspired our simulated datasets to capture key characteristics: (i) a single clinical presentation per trial, (ii) trial-specific baseline hazard rates, and (iii) an association between clinical presentation and endpoint. 

\subsection{Simulated Data}
Our simulated trial data considered: (i) within trial assignment to group $A$ or $B$, (ii) the number of pooled trials, (iii) variable baseline hazard rates (frailty) between trials, and (iv) a hazard ratio of $1.0$ (to study type I error) and $2.0$ (to study power) between patients assigned to group $A$ versus group $B$.
We fixed the number of patients studied to $2,000$, generated $1,000$ simulated datasets, and uniformly divided patients into $T$ trials.
For each trial $t$, we either assigned all patients to group $B$ with probability $p$ (unevenness) or all patients to group $A$ with probability $1-p$.
Thus, within each trial, group assignment was a constant (either all $A$ or all $B$).
Assuming constant hazards $(h_{0})$, we drew event and censoring times from an exponential distribution
\begin{equation*}
  p(T = t) \propto  e^{-h_{0} t}
\end{equation*}
with a $15\%$ event rate at $1825$ days, and $25\%$ censoring rate at $1825$ days.

We defined the hazard rates for group $A$ and $B$ as
\begin{align}
  h(B) &= h_{0} e^{ \log(\text{HR}) + s} \nonumber \\
  h(A) &= h_{0} e^{s} \nonumber\\
  s &\sim LN(\nu,\tau), \nonumber
\end{align}
where $h_{0}$ represents the baseline hazard, HR is the assumed hazard ratio between group $A$ and $B$ and $s$ is a trial-specific quantity (one per trial) drawn from a Log-Normal distribution $(LN)$ centered at $\nu$ with standard deviation $\tau$.
Our simulated data: (i) varied the number of pooled trials, $T$, from $3$ to $10$, (ii) unevenly assigned the proportion $(p)$ of patients among all pooled data to either group $A$ or $B$, and (iii) multiplied half of all pooled trials baseline hazard rates by $\nu$ on average (contrasting baseline hazard rates).
Unevenly assigning groups per trial and separating baseline hazard rates was done to mimic our pooled trial data and demonstrate how pooling trials can bias the Cox-proportional hazards model. 

\subsection{Survival Analysis}

We estimated hazard ratios from the simulated data using (i) a stratified cph model (cph-S), (ii) a cph model including trial as a fixed effect (cph-F), (iii) a cph model including trial as a Gamma distributed random effect (cph-G), and (iv) a cph model including trial as a Log-Normal distributed random effect (cph-L).
Each model manages trial effects differently.

The cph-S model breaks the overall baseline hazard rate into separate trial-specific baseline hazard rates. Given the $p$th patient within the $\tau$th trial:
\begin{equation*}
  h_{\tau,p}(t | x,\beta) = h_{\tau,0}(t) \times  g(x,\beta)
\end{equation*}
describes a trial specific baseline hazard rate $\l(h_{\tau,0}(t)\r)$ and patient specific function $\l(g(x,\beta)\r)$ that depends on patient $p$'s set of covariates $(x)$ and population parameters $(\beta)$.
By separating trials into strata, the cph-S model copes with non-proportional hazards across trials but loses any trial with a single level of the effect of interest.

The cph-F model adjusts for the effect of interest and trial enrollment while assuming patients follow equal hazards through time.
Mathematically,
\begin{equation*}
  h_{p}(t | x,\beta) = h_{0}(t) \times  g(x_{\tau,p},\beta) 
\end{equation*}
where $x$ includes effects for trial and covariates for the patient.
The cph-F model infers hazard ratios from all available data, but assumes proportional hazards between trials.
Compared to stratified models, it trades non-proportional hazards for less selection bias.

The cph-G and cph-L models suppose differences in hazard rates across trial follow a distribution.
Let 
\begin{align*}
  h_{s,p}(t | x,\beta) &= h_{0}(t) \times \phi_{s} \times g(x,\beta)
\end{align*}
where the trial specific effect $\phi_{s}$ (randomly drawn from the Gamma or Log-normal distribution) multiplies each patient's hazards rate.
We multiply patient hazards by random draws from $\phi$ to govern patient differences within trial.

Stratifying, adjusting as a fixed effect, and introducing a random effect for trial represent the three most common paradigms to handle pooled trial data.

\subsection{Statistical inference}

From our 17-trial pooled database, we estimated baseline hazard rate's posterior probability $(h)$ given follow-up times $(d)$ and events $(e)$ as 
\begin{equation*}
  p(h|e,d) \propto h^{\sum{e} + \alpha} \times e^{-\l(\sum{d} + \gamma\r)}
\end{equation*}
considering an exponential model with gamma prior generating the time to event data, and uninformative Gamma prior parameters $\alpha = 10^{-5}$ and $\gamma = 10^{-5}$.

We related any two variables using linear regression with non-informative priors for the intercept $(b)$, slope $(m)$, and variance $(\sigma^2)$.
Mathematically, we relate two variables $V$ and $H$ by
\begin{equation*}
  p(V=v| H=h) \sim  \mathcal{N}\l( b + m \cdot h, \sigma^{2}\r)
\end{equation*}
and compute posterior probabilities for $b$ and $m$ assuming a Normally distributed $\mathcal{N}\l(0,10^{-5}\r)$ prior probability and assuming a Gamma distributed $\mathcal{G}( 4 \times 10^{-2}, 4 \times 10^{-2})$ prior probability for $\sigma^2$.

We compared statistics between two models by fitting a polynomial and averaging over the number of trials $(T)$, unevenness $(p)$, or baseline hazard multiplier $(\nu)$ as
\begin{equation*}
  \bar{S} = \dfrac{1}{r-q}\int_{v=q}^{v=r} f(v) \; dv
\end{equation*}
where $f(v)$ represents a linear or quadratic model, and reported the probability $(\rho)$ any two models differ by more than a relative $\ell\%$ with respect to $R$, or mathematically,
\begin{equation*}
  \rho_{R,\ell} = p \left( \left| \f{X-Y}{R} \right| < \ell \right).
\end{equation*}
We also considered absolute differences between two groups ($X$ and $Y$) as
\begin{equation*}
  \delta_{\ell} = p \left( \left| X - Y \right| < \ell \right).
\end{equation*}
We designated $\rho_{R}$ at $\ell=5\%$, $\delta$ at $\ell=1\%$ unless stated otherwise, and considered $\rho$ or $\delta$ values $<5\%$ significant.

\section{Results}

\begin{figure}[ht!]
  \centering
  \includegraphics[width=\columnwidth]{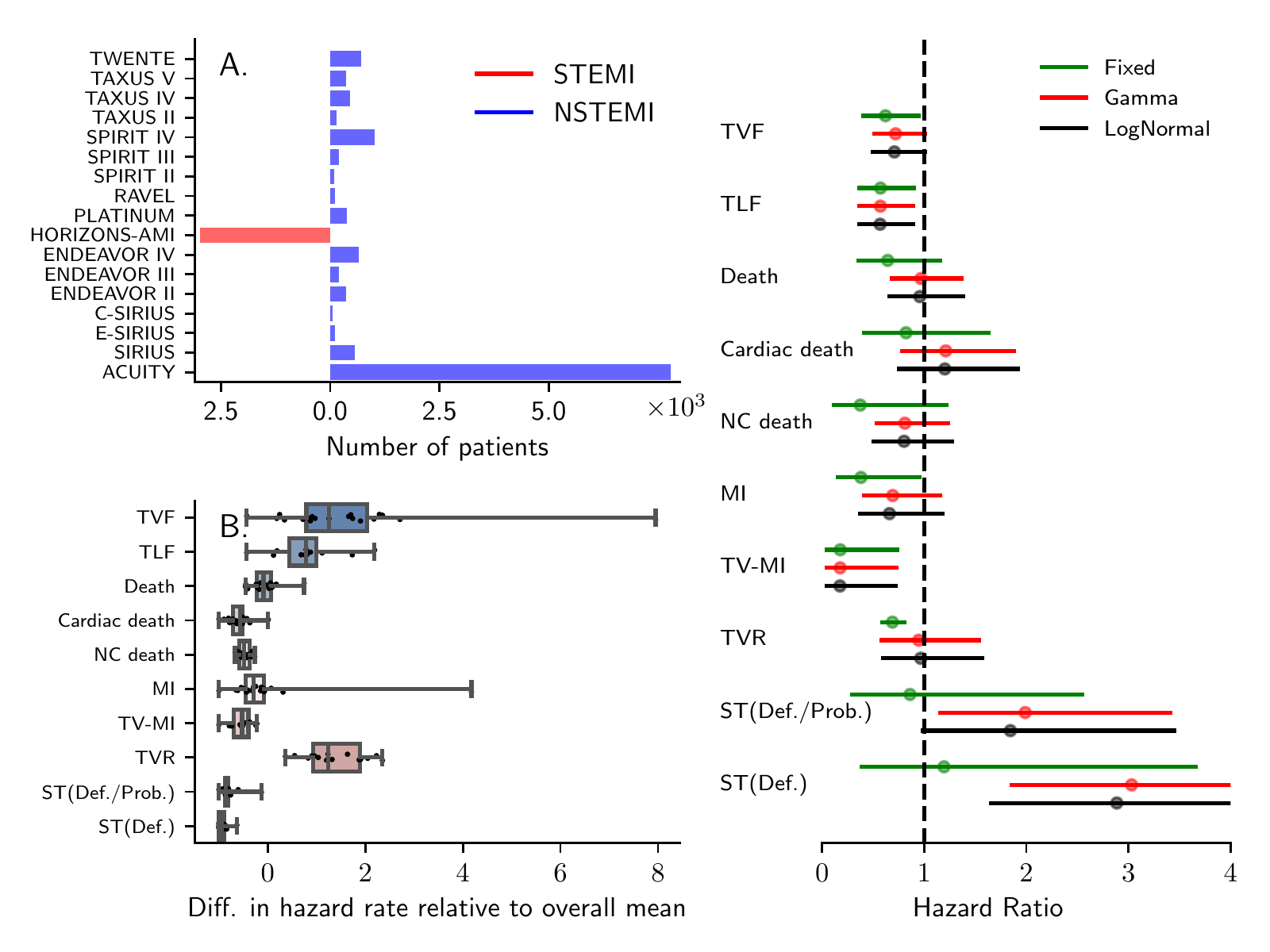}
  \caption{
    After collecting $5$-year event rates from $17$ cardiovascular stent trials (A) we found a strong imbalance in clinical presentation, variable baseline hazard rates for $10$ different clinical endpoints at $5$ years (B), and differences between fixed and random effects Cox proportional hazards models (right).
    (A) Among $17$ clinical trials, only $1/17$ or $5.89\%$ of trials studied STEMI patients.
    (B) We estimated baseline hazard rates using an exponential-gamma model, and reported the relative difference in per-trial baseline hazards compared to the average $(5.15 \times 10^{-5})$ over all $17$ trials and $10$ events.
    The majority of event's baseline hazards noticeably fluctuated by trial.
    (Right) Fixed effect and random effect Cox proportional hazards reported different levels of point and interval estimates and likely related to the imbalance among clinical presentation and variable baseline hazards by trial.
    \label{fig1.compareModels}}
\end{figure}

\begin{figure}[ht!]
  \centering
  \includegraphics[width=\columnwidth]{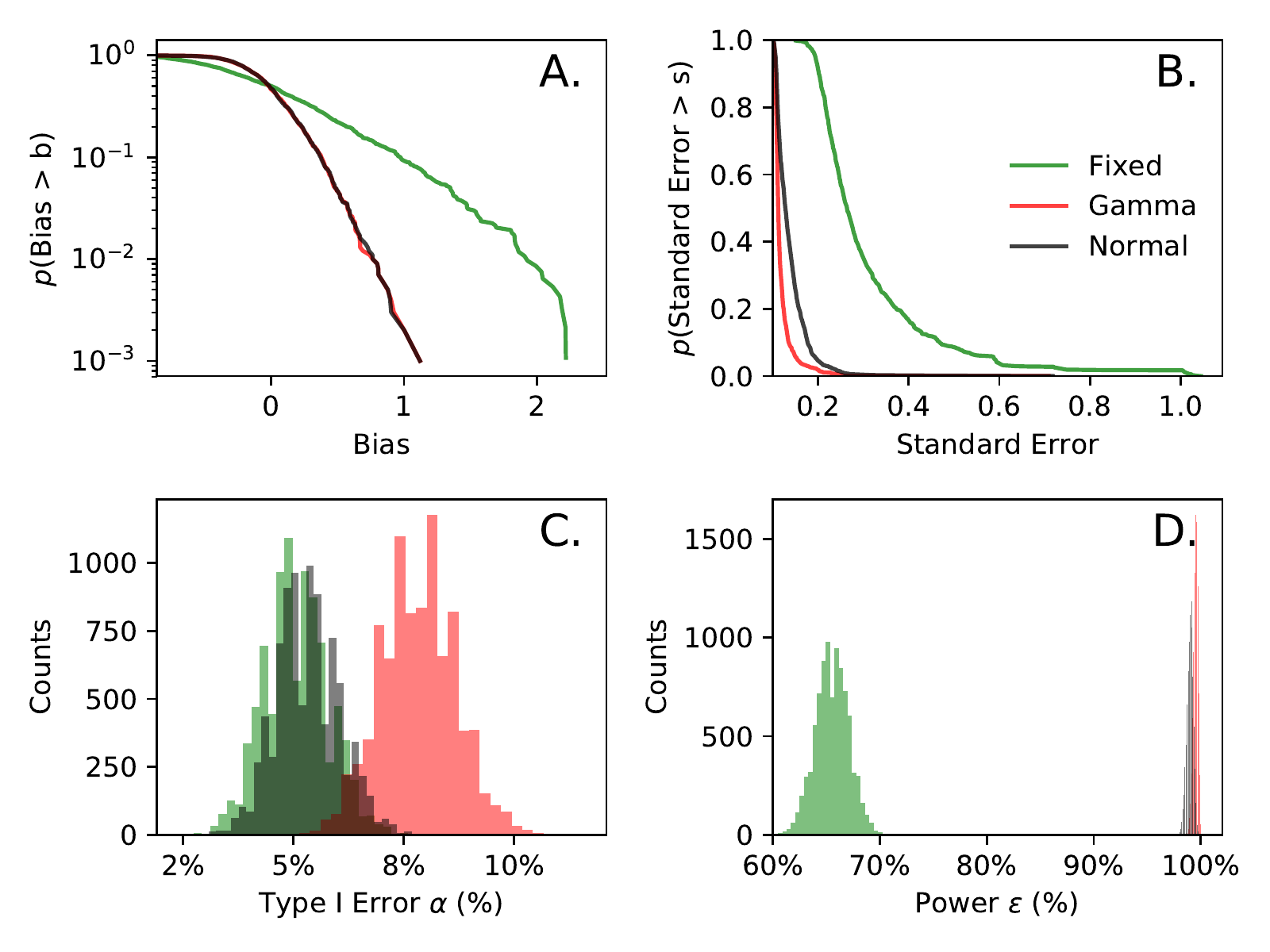}
  \caption{
    For $T = 10$ trials and an even number of trials assigned to group A versus group B $(p = 1/2)$, random effects models had lower bias (A), smaller standard errors (B), variable type I error (C), and superior power (D).
    Random effects models balanced bias and standard error better than fixed models, and this balance resulted in more powerful inference.
    \label{fig2.fixedT}}
\end{figure}

\begin{figure}[ht!]
  \centering
  \includegraphics[width=\columnwidth]{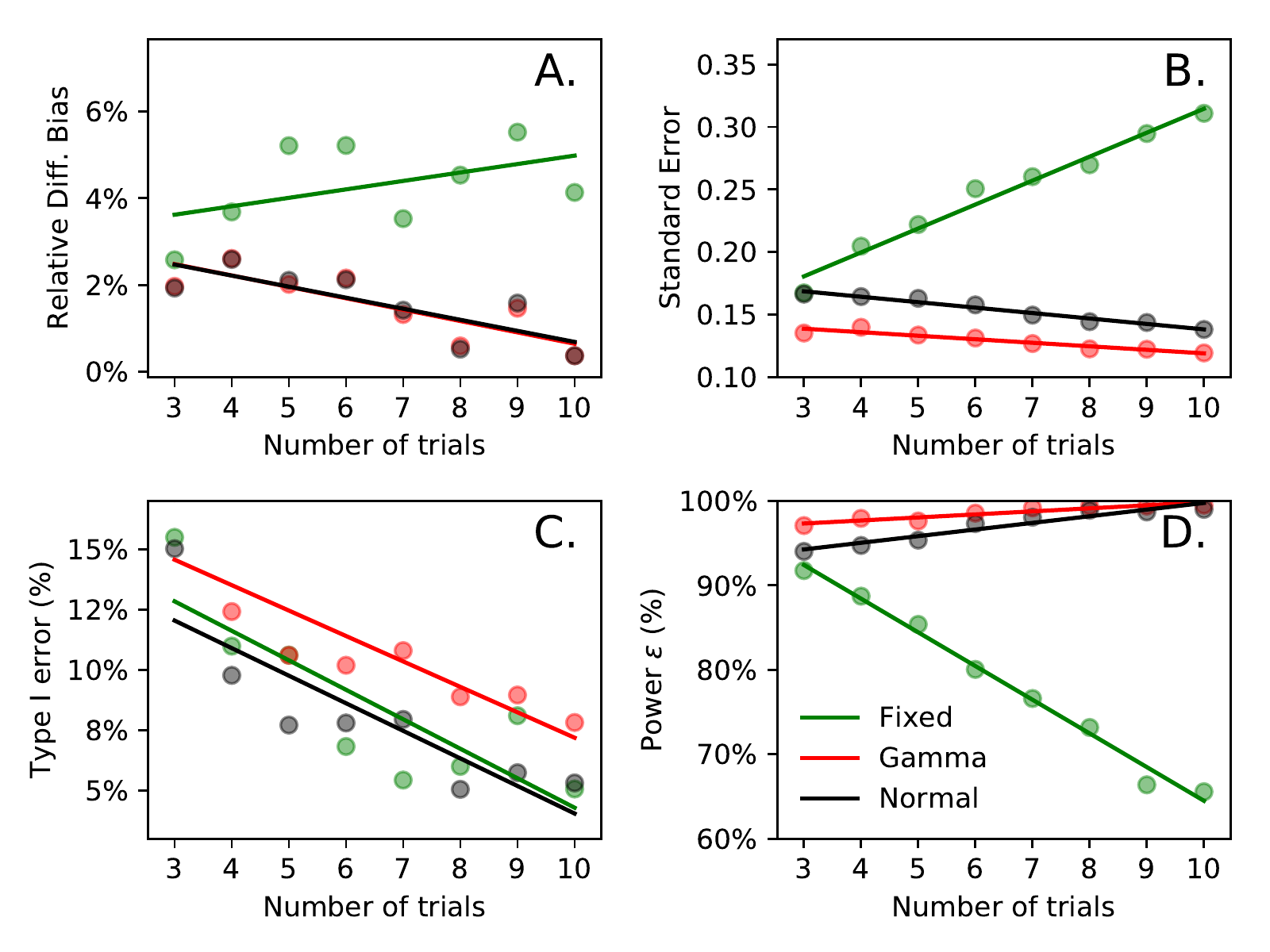}
  \caption{
    Varying the number of pooled trials from $3$ to $10$, random effects models maintained lower bias, standard error, and type I error compared to fixed models.
    Random effects models also had higher power than fixed models.
    An increasing number of trials also raised fixed model bias and standard error while doing the opposite for random effects models; it shrunk type I error for both fixed and random models.
    Pooling more trials also raised the power of random effects models while decreasing fixed models' power.
    Combining more trials intensified uneven patient assignment, breaking the fixed model and strengthening random effects models.
    \label{fig3.ByT}}
\end{figure}

\begin{figure}[ht!]
  \centering
  \includegraphics[width=\columnwidth]{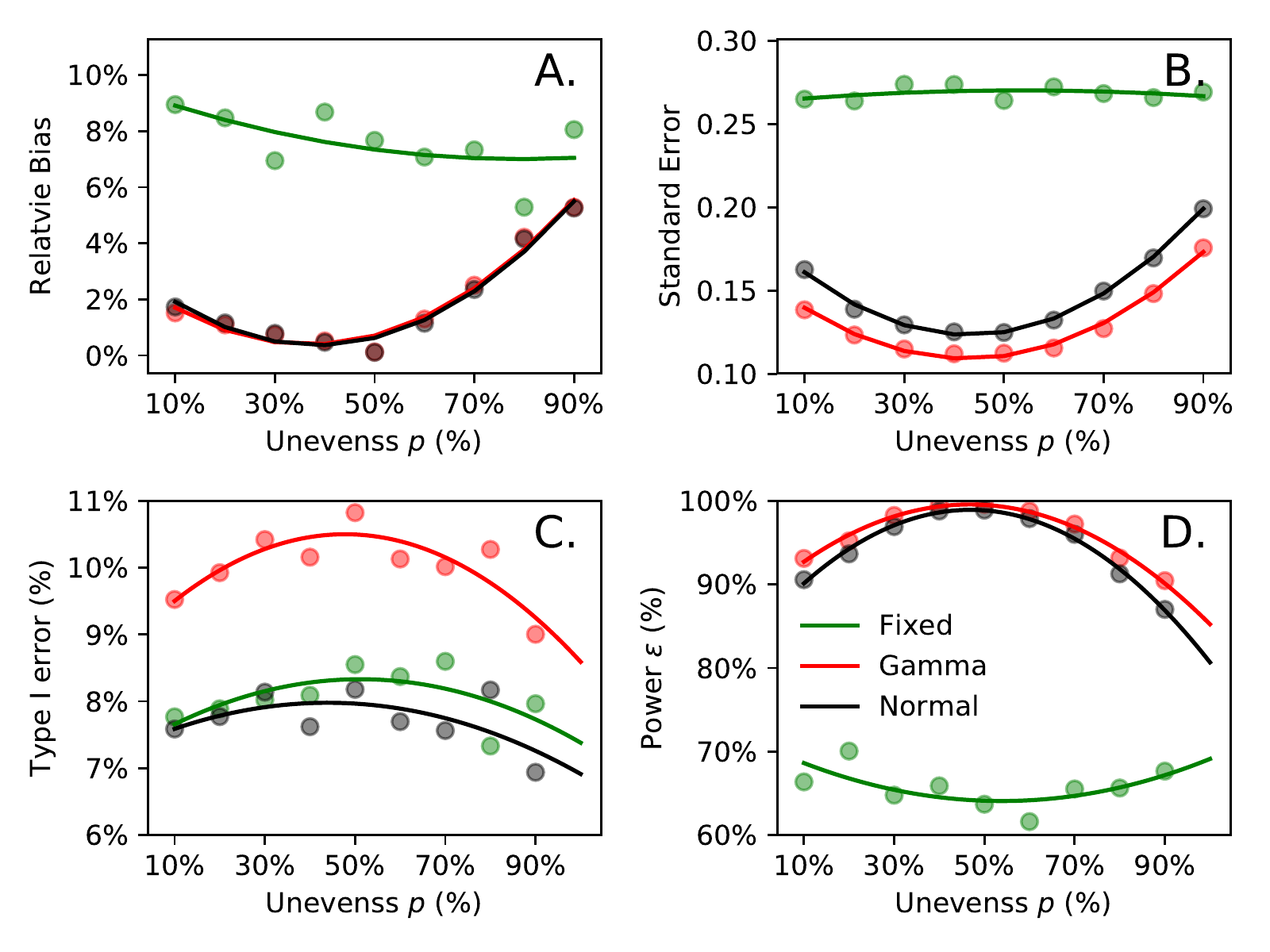}
  \caption{
    When each trial has a single group ($A$ or $B$), all models performed best with an even proportion of group $A$ and $B$ patients across all trials.
    Random effects models had the smallest relative bias (A), lower standard error (B), variable type I error (C), and stronger power (D).
    We observed no difference in relative bias between normal and gamma models, but the gamma model had smaller standard error, inflated type I error, and higher power than the normal model.
    Uneven $(p)$ pooling biases hazard ratio estimates and decreased power in a pooled study.
    \label{fig4.Uneven}}
\end{figure}

\begin{figure}[ht!]
  \centering
  \includegraphics[width=\columnwidth]{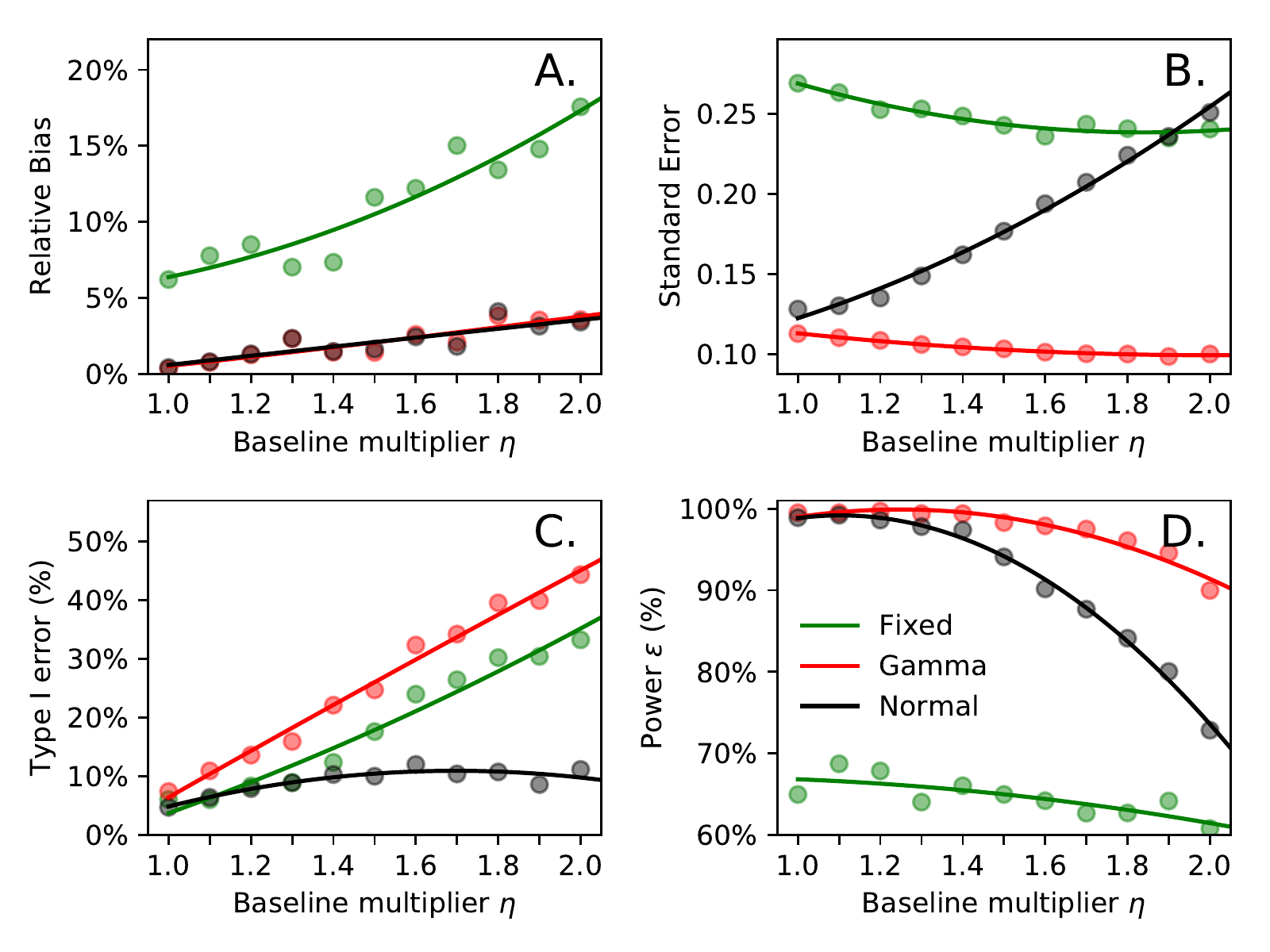}
  \caption{
    Separating between-trial baseline hazard rates inflated random effect models' standard error (B) and decreased power (D).
    While differing baseline hazard rates damaged random effects model power, these models maintained a small bias (A) compared to fixed models.
    We saw heightened biases translated to inflated type I errors in both fixed and gamma models (C).
    Dissimilar baseline hazard rates damaged random effects models.
    \label{fig5.trialBaselineHazard}}
\end{figure}

When studying a trial-specific covariate inspired from real data (Figure~\ref{fig1.compareModels}) with skewed patient assignment across studies and similar baseline hazard rates, random effects models showed less bias, smaller standard error, variable type I error, and increased power over fixed effects models (Figure~\ref{fig2.fixedT}, Figure~\ref{fig3.ByT}, and Figure~\ref{fig4.Uneven}).
Exploring assorted baseline hazard rates across trials, we found both fixed and random effects models suffered (Figure~\ref{fig5.trialBaselineHazard}).
Assigning a single group (all $A$ or all $B$) per trial prevents us from using any model stratified by trial, decreases power and biases fixed models, and random effects models likely traded inflated type I error for stronger hazard ratios and statistical confidence.

We found a heavy imbalance between STEMI/NSTEMI patients~(Figure~\ref{fig1.compareModels}A) and variable baseline hazard rates~(Figure~\ref{fig1.compareModels}B) among $10$ different clinical endpoints across $17$ STEMI/NSTEMI stent trials.
This imbalance and variable baseline hazards likely caused disagreement between the fixed and random effects models (Figure~\ref{fig1.compareModels} Right).
Our pooled data set contained one trial studying STEMI patients (HORIZONS-AMI) and $16$ trials studying NSTEMI patients.
After estimating baseline hazard rates for all $10$ endpoints among $17$ trials, we found an overall mean equal to $5.15 \times 10^{-5}$ and relative trial differences from $-1.00$ to $7.95$ times the mean.
Studying an imbalanced STEMI/NSTEMI covariate with variable baseline hazards led to large disagreement between fixed and random effect models for: death, cardiac death, non-cardiac death, myocardial infarction, target vessel revascularization, and stent thrombosis.
Our simulated data aimed to tease apart why we found these disagreements between models.

For $T=10$ trials~(Figure~\ref{fig2.fixedT}), the fixed effects model biased hazard ratio estimates more ($\rho_{\mathrm{Ran.}}<0.001$, Figure~\ref{fig2.fixedT}A), had inflated standard errors ($\rho_{\mathrm{Ran.}}<0.001$, Figure~\ref{fig2.fixedT}B.), comparable type I error ($\rho_{\mathrm{Ran.}} = 0.069$, Figure~\ref{fig2.fixedT}C.), and diminished power ($\rho_{\mathrm{Ran.}} < 0.001$, Figure 1D) compared to random effects models.
The gamma versus normal random effects models had comparable bias $(\rho_{\mathrm{Normal}, 0.5}=1.0)$, smaller standard error ($\rho_{\mathrm{Normal}} = 0.044$), elevated type I error $(\rho_{\mathrm{Normal}} = 0.016)$, and improved power $(\rho_{\mathrm{Normal}} < 0.001)$.
The bias did not significantly differ between random effects models, but compared to the normal model, the gamma model's smaller standard error magnified type I error and strengthened power.
We uncovered each model's strengths and weaknesses by studying a static $T=10$ trials.
Differing the number of trials and studying the same model properties helped determine how random effects models compare to fixed effects models under more variable baseline hazard rates.

When varying the number of trials~(Figure~\ref{fig3.ByT}) and averaging over simulations, random effects models maintained a lower bias ($\rho_{\mathrm{Fix.}} = 0.026$, Figure~\ref{fig3.ByT}A.), reduced standard error ($\rho_{\mathrm{Fix.}} = 0.039$, Figure~\ref{fig3.ByT}B.), variable type I error ($\rho_{\mathrm{Fix.}} <0.001$, Figure~\ref{fig3.ByT}C.), and improved power ($\rho_{\mathrm{Fix.}} <0.001$, Figure~\ref{fig3.ByT}D.) compared to fixed effect models.
The fixed effect model biased hazard ratios more than random effects models and resulted in elevated type I errors, but inflated standard errors robbed the fixed model of power.
A similar power/type I error tradeoff persisted between random effects models when varying the number of trials.
The gamma model had smaller standard error than the normal model ($\rho_{\mathrm{Normal}}=0.001$), more type I error than the normal model ($\rho_{\mathrm{Normal}} = 0.027$), and similar power to the normal model ($\rho_{\mathrm{Normal}}=1.0$).

Pooling studies with an unequal ratio of group $A$ to $B$ (increasing unevenness,~Figure~\ref{fig4.Uneven}) heightened bias ($\delta = 0.034$, Figure~\ref{fig4.Uneven}A), inflated standard error ($\delta = 0.018$, Figure~\ref{fig4.Uneven}B), reduced type I error ($\delta = 0.034$, Figure~\ref{fig4.Uneven}C), and reduced power in all models ($\delta<0.001$, Figure~\ref{fig4.Uneven}D) when averaging over simulations.
Fixed models biased hazard ratio estimates $2.8$ times more than random effects models ($\rho_{\mathrm{Normal}} = 0.026$), inflated standard errors $1.14$ times more than random effects models ($\rho_{\mathrm{Normal}}=0.11$), and lowered power $1.96$ times more than random effects models ($\rho_{\mathrm{Normal}} <0.001$).
Compared to fixed models, the normal model shrunk type I error $1.56$ times ($\rho_{\mathrm{Normal}} = 0.22$) but the gamma model raised type I error $1.76$ times ($\rho_{\mathrm{Normal}} = 0.97$).
Random effects models better managed bias, standard error, and power, but inflated type I error compared to fixed models.

Separating baseline hazards between trials (Figure~\ref{fig5.trialBaselineHazard}) inflated type I error ($\delta<0.001$, Figure~\ref{fig5.trialBaselineHazard}C) and weakened power ($\delta<0.001$, Figure~\ref{fig5.trialBaselineHazard}D) in random effect models after averaging over simulations.
Fixed models sustained increasing bias ($\delta<0.001$, Figure~\ref{fig5.trialBaselineHazard}A) and type I error ($\delta<0.001$, Figure~\ref{fig5.trialBaselineHazard}C).
Fixed models also suffered from large standard error ($\delta<0.001$, Figure~\ref{fig5.trialBaselineHazard}B) and low power ($\delta<0.001$), but increasing between-trial baseline hazard rates escalated the normal model's standard error ($m \pm \mathrm{std.err.} = 0.062 \pm 0.225$) and decreased power ($m \pm \mathrm{std.err.} = -12.595 \pm 1.512$).
Diverse baseline hazards also inflated the gamma model's type I error ($m \pm \mathrm{std.err.} = 19.337 \pm 0.773$).
Contrasting baseline hazard rates between trials damaged random effects models.

Although unevenly distributing patient groups across trials damaged hazard ratio estimates, random effects models stayed unbiased and maintained power compared to stratified and fixed effect models.
Disparate baseline hazards damaged random effects models.
We saw our simulated results replicated in a $17$-trial dataset; random effects models performed best when confronted with an unevenly distributed covariate.

\section{Discussion}

Our simulated experiment showed both fixed and random effects models fail when studying a covariate unevenly distribute across trials that have varying baseline hazard rates.
We applied fixed and random effect models to a real $17$-trial dataset of STEMI and NSTEMI patients and found worse performance for fixed effect models than random effects models when trials had similar baseline hazard rates.
Unlike previous studies that scrutinize cluster variability, this work tests model robustness under a fixed-effect's imbalance across clusters.

Clinical science strives to better understand scarce disease types, but these scarce disease types and trial diversity will intensify survival rate variability.
When we pool clinical data to study rare events, we need to collect comprehensive outcomes data from many trials or resort to more robust models.
This increased survival variability may result in contrasting baseline hazard rates.

After comparing fixed effect and random effects models, we found random effects models performed better in bias, standard error, and power, but failed when pooling trials with dissimilar baseline hazards.
We speculate the difficulty estimating a single set of parameters for dissimilar baseline hazards causes random effects models to fail.
When pooling trials with contrasting baseline hazards, including a covariate that groups trials by baseline hazards may help.

Compared to fixed models, random effects models take better advantage of trial information to reduce hazard ratio estimates' standard errors.
Stratified models estimate treatment effect within each trial, and with only one treatment per trial, have no use.
Fixed effect models do take advantage of treatment effect across all trials, but estimating effects for each trial creates a more uncertain treatment-effects hazard ratio.
Random effects models isolate treatment assignment (as a fixed effect) from trial effects (as a random effect) and apply additional trial data toward better estimating treatment assignment.

We limited ourselves to studying a binary covariate unevenly distributed across trials, simulating data with constant hazards, and modeling time to event with cph models.
A continuous covariate could behave differently than a binary variable under the same conditions.
Constant hazards do not occur in typical trials, but cph models ignore baseline hazard rates.
Many other models exist to model time to event, such as accelerated failure time models, and may prove more useful than cph.
This study's limitations may inspire more realistic simulated datasets and their effect on the cph model.

We plan to study whether Cox proportional hazards models perform better with a smaller set of trials with covariate balance compared to a larger set of unbalanced trials.
We also plan to study continuous variables split across pooled trials and alternative models of time-to-event data within the context of unevenly distributed variables across a pooled database.

In most cases, random effects models stand up to an unevenly distributed binary covariate so long as each trial has similar baseline hazard rates.

\section*{Acknowledgments}
The authors thank Karl~Lherisson and the Cardiovascular Research Foundation's Information Technology department for computational resources.

\subsection*{Financial disclosure}

None reported.

\subsection*{Conflict of interest}

The authors declare no potential conflict of interests.


\end{document}